\documentclass[11pt,a4paper]{article}
\pdfoutput=1
\usepackage{graphicx,sectsty,amssymb,amsmath,color}
\usepackage[linkcolor={blue},citecolor={red},colorlinks=true]{hyperref}
\usepackage{enumerate}
\usepackage[margin=2cm]{geometry}
\usepackage{multirow}
\usepackage{authblk}
\usepackage{slashed}

\newcommand{\beq}{\begin{equation}}
\newcommand{\eeq}{\end{equation}}

\newcommand{\cO}{\mathcal{O}}
\newcommand{\cL}{\mathcal{L}}

\newcommand{\BR}{{\rm BR}}

\definecolor{red1}{cmyk}{0,1,1,0.3}

\def\beq{\begin{equation}}
\def\eeq{\end{equation}}
\def\bea{\begin{eqnarray}}
\def\eea{\end{eqnarray}}

\def\bit{\begin{itemize}}
\def\eit{\end{itemize}}

\def\l{\left}
\def\r{\right}

\def\ra{\rightarrow}

\def\baa{\begin{array}}
\def\eaa{\end{array}}
\def\sh2{s_{\frac{h}{2}}}
\def\ch2{c_{\frac{h}{2}}}

\def\sl#1{\mathord{\not\mathrel{{\mathrel{#1}}}}}
\def\d{\partial}

\def\simgt{\mathrel{\lower2.5pt\vbox{\lineskip=0pt\baselineskip=0pt
           \hbox{$>$}\hbox{$\sim$}}}}
\def\simlt{\mathrel{\lower2.5pt\vbox{\lineskip=0pt\baselineskip=0pt
           \hbox{$<$}\hbox{$\sim$}}}}

\def \dblarrow#1{\overset{\leftrightarrow}{#1}}

\begin{document}

\title{
{\small\begin{flushright}
\small{CERN-PH-TH-2014-159}
\end{flushright}}
\vspace{0.5cm}
On the Flavor Structure of
Natural Composite Higgs Models \& Top Flavor Violation}
\date{}
\author[1]{\small Aleksandr Azatov\thanks{aleksandr.azatov@cern.ch}}
\author[1]{\small Giuliano Panico\thanks{giuliano.panico@cern.ch}}
\author[1,2]{\small Gilad Perez\thanks{gilad.perez@weizmann.ac.il}}
\author[2]{\small Yotam Soreq\thanks{yotam.soreq@weizmann.ac.il}}

\affil[1]{\it\small CERN Physics Department, Theory Division, CH-1211 Geneva 23, Switzerland}
\affil[2]{\it\small Department of Particle Physics and Astrophysics, Weizmann Institute of Science, Rehovot 7610001, Israel}

\maketitle

\begin{abstract}
We explore the up flavor structure of composite pseudo Nambu-Goldstone-boson Higgs models, where we focus on the flavor anarchic minimal $SO(5)$ case.
We identify the different sources of flavor violation in this framework and emphasise the differences from the anarchic Randall-Sundrum scenario.
In particular, the fact that the flavor symmetry does not commute with the symmetries that stabilize the Higgs potential may constrain the flavor structure of the theory. In addition, we consider the interplay between the fine tuning of the model and flavor violation. We find that generically the tuning of this class of models is worsen in the anarchic case
 due to the contributions from the additional fermion resonances.
 We show that, even in the presence of custodial symmetry, large top flavor violating rate are naturally expected.
 In particular, $t\to cZ$ branching ratio of order of $10^{-5}$ is generic for this class of models.  Thus, this framework can be tested in the next run of the LHC as well as in other future colliders.
 We also find that the top flavor violation is weakly correlated with the increase amount of fine tuning.
Finally, other related flavor violation effects, such as $t \to ch$ and in the $D$
system, are found to be too small to be observed by the current and near future colliders.
\end{abstract}

\section{Introduction} \label{sec:intro}

Now that the Higgs field has been discovered the standard model~(SM) with its minimal scalar sector of electroweak~(EW) symmetry breaking is
complete. Though at first glance it appears as a rather baroque theory it actually possesses a rather unique structure. If one considers only marginal operators the SM admits a set of accidental (exact and approximate) symmetries leading to: baryon-lepton conservation, suppression of processes involving flavor changing neutral currents~(FCNC) and CP~violation, and special relations among its EW parameters.
Several decades of experimental effort failed to find any flaw in this picture. Furthermore, viewing the SM as an effective description of nature, one
finds that the leading deformation of its structure is expected to induce (Majorana) neutrino masses and mixing. The common explanation for the
neutrino oscillation is that at least two of them are massive, without a clear flavor structure. Alas, the well-know See-Saw mechanism points to an
extremely high scale. A scale way beyond the reach of future experiments and also too high to lead to visible effects in flavor or EW precision measurements.

The above tremendous success of the SM might suggest that our microscopic world is of a form of a ``desert", namely it does not contain any kind
of new dynamics for very many decades of energy scales. However, there is one basic question that is left unanswered within the SM, that can be
associated with a nearby energy scale: how come the SM spectrum contains a light fundamental scalar?
As is well known the masses of scalar fields are radiatively unstable, this implies that Nature is finely tuned, unless a new form of physics that is
coupled to the SM Higgs and EW sector exists at the TeV scale.
An effective minimal approach to naturalness could be realized by extending different SM sectors according to their ``naturalness" pressure.
From low energy perspective one can order the SM different fields according to the size of their individual radiative contributions to the Higgs
mass square.
The largest contribution are the ones that are associated with the Higgs large coupling to the top quark.  In such a way the sector with the lowest
new physics~(NP) scale, possibly beside that of the Higgs itself, would be the top sector.
In the effective theory the NP contributions to various SM rare processes are inversely proportional to the NP size. Thus, if the above ``inverted"
hierarchical NP scales can be indeed realized in a microscopic theory then the resulting theory would approximately posses the same set
of accidental symmetries as the SM. Such a framework could lead to a viable phenomenological description at least at zeroth order.

The above pattern is by definition non-universal in flavor space, as the sector that contains the light generation quarks would correspond to a much
higher NP scale than the one related to the top sector. When the NP dynamics involves non-universal flavor couplings it would
generically lead to flavor violation. The size of the NP contributions to FCNCs would be proportional to the level of misalignment between
the NP coupling and the corresponding SM Yukawa couplings. One can adopt the approach in which the NP couplings follow
exactly the same pattern as the SM ones, that is dictated by the Yukawa interactions. This possibility is denoted as minimal flavor violation~(MFV)~
(see~\cite{D'Ambrosio:2002ex} and also~\cite{Kagan:2009bn} for symmetry based descriptions). However, such an approach suffers from two conceptual drawbacks. The first is that MFV is just an
ansatz. It does not shed light on the flavor puzzle, namely why the quark flavor parameters are small and hierarchical. The second, which is
probably more relevant for TeV physics is that accommodating the MFV ansatz makes the NP structure highly non-generic and imposes a
serious burden in terms of model building. Possibly a more generic approach would be one in which there is some semi-universal parametric
power counting that controls the strength of the interaction between the different SM matter fields and the Higgs  and their coupling to the new
physics sector. Within such a framework one can obtain a unified explanation for the SM flavor puzzle and a protection against overly large
contributions to various FCNC processes. However, as the suppression of the various coupling in the theory is only parametric and not exact we do
expect contributions to flavor changing processes to arise at some level.

Focusing below on the quark sector we can identify two sources of parametric suppression: one is related to ratio of masses and the other is
related to ratio of mixing angles and associated with the structure of the left handed (LH) charged currents. Following the above rationale we thus
expect the ratio between the coupling of the NP degrees of freedom to the SM LH (weak doublets) field to be of the order of the CKM elements and
for the right handed (RH) ones we expect the ratio of couplings to be of the order of the masses divided by the corresponding mixing angles.
Denoting the strength of the relevant interaction between the SM fields and the NP as being proportional to a parameter $\lambda_{L,R}^i
$ ($i=1..3$ is a generation index and $L,R$ is related to doublet and singlet respectively) one finds~(see {\it e.g}~\cite{Isidori:2010kg}),
\bea
&&\lambda^3_L:\lambda^2_L:\lambda^1_L\sim 1:V_{cb}:V_{ub}\sim 1: 4 \times 10^{-2} : 4 \times 10^{-3} \, , \nonumber\\
&&\rule{0pt}{1.5em}\lambda^3_{R,u}:\lambda^2_{R,u}:\lambda^1_{R,u}\sim 1: \frac{m_c}{m_t V_{cb}} :\frac{m_u}{m_t V_{ub}}\sim
1: 9\times 10^{-2}: 2\times10^{-3}  \, ,
 \nonumber\\
&&\rule{0pt}{1.5em}\lambda^3_{R,d}:\lambda^2_{R,d}:\lambda^1_{R,d}\sim 1: \frac{m_s}{m_b V_{cb}} :\frac{m_d}{m_b V_{ub}} \sim 1: 0.4: 0.2  \, ,
\label{eq:flavor_anarchy}
\eea
where the quark masses are taken from~\cite{Xing:2011aa} at $1\,$TeV and the CKM elements from~\cite{Beringer:1900zz}. The relations above
provide us with patterns of flavor violation that can be confronted with data. However, these are generic and they give no information regarding
the overall scale of NP and also what the nature of the NP dynamics is. Nevertheless, assuming a universal scale one can
already identify what the most constraining set of observables in such a framework would be~\cite{Bona:2007vi, Davidson:2007si, Agashe:2008uz,
Gedalia:2009ws, KerenZur:2012fr}. It is quite obvious that the constraints coming from the down sector are stronger than the ones coming from the
up one~\cite{Isidori:2010kg}. Furthermore, as in this paper we are interested to study $Z$-mediated FCNC one should also bear in mind that
EW~precision measurements severely constrain non-SM shifts in the $Z$ to $b\bar b$ coupling. This motives us to look for effects in the up sector
where the constraints are weaker~\cite{Nir:1993mx, Csaki:2009wc,Fitzpatrick:2007sa}. Moreover, examining the above relations it seems obvious
that the first place to be looking for flavor violation is in the top sector, namely in top-charm transitions. Examining~Eq.~\eqref{eq:flavor_anarchy} at
any rate reveals that the largest source of flavor violation would be within the RH sector in transitions related to RH top decaying to RH charm, with
the LH transitions being suppressed only by a factor of few. All this seems to give a pretty strong motivation to study the $t\to c Z$ process.

Within the SM the branching ratio of $t\to cZ$ is highly suppressed and expected to be at the order of $10^{-13}$~\cite{Eilam:1990zc}. Thus, any
signal well above it is a clear signal of NP. Currently, the searches of $t\to cZ$ give null result and set an upper bound of $\BR(t\to
cZ)<5\times10^{-4}$ at 95\%\,CL~\cite{Chatrchyan:2013nwa}. At the LHC this constraint can be improved by an order of magnitude. Notice that,
when considering rates below $\BR(t\to cZ)<5\times10^{-5}$ or so, one needs to carefully take into account SM backgrounds coming
from production of top $Z$ and a jet~\cite{Campbell:2013yla}. It seems thus that a branching ratio of roughly $10^{-5}$ is particularly timely and
relevant when discussing $t\to cZ$. However, given the large enhancement compared to the small SM rate it would be fair to ask whether such a
rate can be expected from any well motivated (reasonable) extension of the SM. To get some perspective on the size of the required effect let us
use simple effective field theory (EFT) to see how large is the expected rate.

The structure of the EFT mediating $(t\to cZ)$ is in fact pretty simple. Only three dimension-six operators are relevant for our discussion~\cite{Fox:2007in, CorderoCid:2004vi,delAguila:2000aa,AguilarSaavedra:2008zc,Grzadkowski:2010es}:
\begin{align} \label{eq:dim6}
	\bar t_R \gamma^\mu c_R (H^\dagger \dblarrow{D}_\mu H) \, , \qquad
	\bar t_L \gamma^\mu c_L (H^\dagger \dblarrow{D}_\mu H)  \qquad \mathrm{and} \qquad
	\bar t_L \gamma^\mu\sigma_3 c_L (H^\dagger \sigma_3\dblarrow{D}_\mu H) \, .
\end{align}
In order to fix our notation we write the generic flavor violating terms in the Lagrangian involving the top and charm quarks and the $Z$ boson as
\begin{align}
	\cL^{tc}_{\rm int}
=	\left(g_{tc,L} \bar{t}_L \gamma^\mu c_L + g_{tc,R} \bar{t}_R\gamma^\mu c_R \right)Z_\mu \, + h.c.\,.\label{basis}
\end{align}
The branching fraction for the decay of the top into the charm quark and the $Z$ can be expressed in terms of the $g_{tc,L}$ and $g_{tc,R}$
couplings and is approximately equal to 
\begin{align}
	\BR(t\to c Z) &\simeq 3.5\times \left(g^2_{tc,R}\, ,\, g^2_{tc,L}\right)
	\approx 3.5 \left(\frac{g}{2 c_W}\right)^2 \l(\left(\frac{m_c  }{m_t V_{cb}}\right)^2\,,\,V_{cb}^2\r)\frac{v^4}{4M_*^4}\nonumber  \\
	&\sim \left(1.5\,,\, 0.3\right) \times 10^{-5} \,\l(\frac{700\,\rm GeV }{M_*}\r)^4\, .\label{generic}
\end{align}
 where $v=246\,$GeV and $c_W$ is the cosine of the Weinberg angle. We have written Eq.~\eqref{generic} in a way that reflects the various
parametric suppression that control the flavor violation of our theory  as explained above. We have also use $M_*$ to describe the scale that
controls the ``microscopic" scale of our effective theory. We can learn several interesting things by examining Eq.~\eqref{generic} as follows.
\begin{itemize}
\item We find that a rather low effective scale is required to have $\BR(t\to c Z)$ of the order of $10^{-5}$. However, such a scale is in fact motivated
by naturalness. Indeed, it is roughly the scale where we expect new degrees of freedom related to the extended top sector to be present.
\item The small scale required further implies that theories in which top FCNC arises at the one-loop level, or that are controlled by weak couplings
are probably out of the reach of the current and next run of the LHC. Furthermore, in this case the expected low rates implies that the search would
not be background free anymore.
\item We find that we expect that the contributions to RH flavor violating currents would dominate over the LH ones. In case of an observed signal
it would be actually straightforward to test this prediction.
As the searches target $t\bar t$ events that are produced via QCD hence both tops are expected to be of the same chirality and one can use the
standard top polarization tests to study the polarization of the flavor violating top as well as the top on the ``other side" of the event.
\end{itemize}

The above serves as a motivation to study top FCNC in models of strong dynamics where the relevant couplings are expected to be sizeable and
flavor violation arises at tree level. In~\cite{Agashe:2006wa} such a scenario was analyzed in the context of anarchic RS models, or their dual
composite Higgs models with partial compositeness, in which the parametric suppression of Eq.~\eqref{eq:flavor_anarchy} holds.
The above qualitative results described in the bullets where indeed confirmed and $\BR(t\to cZ)\sim10^{-5}$ was generically expected for Kaluza--Klein scale of roughly 3\,TeV and a RH coupling of roughly 5 times that of the $Z$, consistent with the scales given in Eq.~\eqref{generic}.
The LH coupling was assumed to be suppressed even beyond the general suppression found in~\eqref{generic} as follows.
The theory that has been considered in~\cite{Agashe:2006wa} was not phenomenologically viable as it predicted a too large shift in the coupling of
$Z$ to $b\bar b$. The large top mass is forcing the LH top doublet to be of a minimal level of compositeness that leads to this overly large
non-oblique shift in the $Zb_L \bar b_L$ couplings (and at the same time further reduce the LH contributions to $\BR(t\to cZ)$).
Furthermore, the model discussed in~\cite{Agashe:2006wa} suffered from a severe little hierarchy problem as the Higgs boson was not realized as
a pseudo Nambu-Goldstone boson (pNGB).

A custodial symmetry to protect the $Zb_L \bar b_L$ coupling against the above overly large contributions was introduced in~\cite{Agashe:2006at}
and a theory with composite LH tops and bottom was becoming phenomenologically viable. However, at the same time, in custodially protected
models it is rather generic to obtain protection of the $Z$ coupling to RH tops and the generic prediction of Eq.~\eqref{generic} for the RH currents
is lost. As the custodial symmetry can only protect one component of a custodial multiplet the $Z \bar t_L t_L$ coupling might be significantly shifted
from its SM value. This naively lead to rates dominated by the LH current that are however further suppressed according to~\eqref{generic}.

In this work we consider the $t\to c Z$ process in composite models where the Higgs is pNGB and as a result the little hierarchy problem is
ameliorated. As we discuss below, it leads to several quite generic consequences that basically resurrect this process as an important probe of this
framework:
\begin{itemize}
\item[(i)] In pNGB composite Higgs models, the physical elementary--composite mixing parameter is effectively an angle and thus constraint to be
of order unity at most. It implies that to accommodate the large top Yukawa both the LH and RH top component need to be sizable.
\item[(ii)] As already mentioned, sizable LH compositeness implies LH top FCNC though the rate is somewhat suppressed as it is proportional to
$V_{cb}$\,.
\item[(iii)] Minimizing the tuning of the class of pNGB models studied by us typically implies the need to maximise the level of top LH
compositeness. This, in fact implies that through left-right mixing the presence of LH custodial violation leads to a larger $\BR(t\to cZ)$ through the
RH large size of composite flavor violation.
\item[(iv)] Finally, we point that, in anarchic pNGB models the source of flavor violation is obtained through the misalignment between the partial
composite mixing matrices and the mass matrix of the vector-like composite fermonic partners.
It implies that the would be top partner is not a pure mass eigenstate which generically leads to worsening of the level of fine tuning of this
framework. It is a manifestation of a rather rare semi-direct linkage between the physics of naturalness of that of flavor violation. (For a related
discussion see~\cite{Blanke:2013uia,Delaunay:2013pwa})
\end{itemize}

In the following section we introduce the set-up and discuss its flavor structure and the tuning in it. Section~\ref{sec:MCHM5} contains the details of
top flavor violation. In Section~\ref{sec:Htc}, we discuss Higgs flavor violation, while in Section~\ref{sec:Dphysics} relevant effects in $D$ physics
are discussed. We conclude in Section~\ref{sec:Conclusions}. Appendix~\ref{app:diagM} contains relevant details for the diagonalization of the up
type quarks mass matrix.

\section{The set-up} \label{sec:setup}

In this work we will consider the simplest realization of the composite Higgs scenario
in which the Higgs boson arises as a Nambu--Goldstone boson of the spontaneously
broken global symmetry $SO(5)/SO(4)$. For definiteness we will assume that the elementary
SM fermions are linearly mixed with composite operators in the fundamental representation
(the $\mathbf 5$) of the global $SO(5)$ group. This framework realizes the partial compositeness mechanism~\cite{Kaplan:1991dc} for the
generation of the SM fermion masses and is usually known
in the literature as the MCHM$_5$ set-up~\cite{Contino:2006qr}. Notice that this is
the minimal composite Higgs implementation that includes a custodial $P_{LR}$ protection
for the bottom coupling to the $Z$ boson~\cite{Agashe:2006at}.

In our analysis we will use an effective field theory approach and we will describe
the low-energy dynamics of the theory by the use of the Callan--Coleman--Wess--Zumino
formalism~(CCWZ)~\cite{CCWZ1}, which allows to write the most general Lagrangian compatible
with the Goldstone nature of the Higgs. In addition to the Higgs boson and the elementary
components of the SM fields, our effective description also contains a set of composite resonances
coming from the strong-sector dynamics. For simplicity, we parametrize the composite resonances
(in particular the fermionc ones that are relevant for our analysis) by the so called
``two site'' construction~\cite{Contino:2006nn, Panico:2011pw, DeCurtis:2011yx,
Matsedonskyi:2012ym,Carena:2014ria}, which corresponds to parametrizing the composite sector by just one level of composite fields. Notice that this simplification does not spoil the main features of the composite Higgs scenarios,
namely the Goldstone nature of the Higgs and the calculability of the model (in particular of the Higgs potential). Moreover it retains the lightest
composite resonances that generate the leading contributions to the flavor effects we are investigating.

The goal of this study is to analyze the flavor violating interactions of the $Z$ boson with the
top and charm quarks as well as the flavor structure of the theory. This allows us to simplify the discussion by ignoring completely
the first generation of the SM fermions and their composite partners. The inclusion of these effects,
in fact, would not generate any qualitative change in our results. However, these effects will be considered in Section~\ref{sec:Dphysics}, where we
discuss $D$ physics.

Let us now describe the effective model that is considered below. As we said before, the elementary states corresponding to the second ($q_L^c$ and $c_R$) and third ($q_L^t$ and $t_R$) SM quark generations are mixed with composite operators transforming in the fundamental representation of $SO(5)$. Although the elementary fermions do not fill complete $SO(5)$ representations it is useful to formally restore the global invariance by incorporating them as incomplete multiplets:
\begin{align} \label{eq:qtdef}
	q^{t}_L
=	\frac{1}{\sqrt{2}}
	\begin{pmatrix}
		b_L \\ -ib_L \\ t_L \\ it_L \\ 0
	\end{pmatrix} \, , \qquad
	u^{t}_R
=	\begin{pmatrix}
		0 \\ 0 \\ 0 \\ 0 \\ t_R
	\end{pmatrix} \, .
\end{align}
Analogous embeddings are used for the second generation quarks ($q_L^c$ and $u_R^c$). For convenience, we also represent the composite states as fields filling multiplets in the fundamental representation, $\bf{5}$. Notice, however, that in the CCWZ formalism the composite states transform only under $SO(4)$ transformations, thus, in the most general Lagrangian each $SO(4)$ multiplet can be thought as an independent component. The explicit form of the $SO(5)$ composite multiplets is
\begin{align}
	\psi^i
=	\frac{1}{\sqrt{2}}
	\begin{pmatrix}
		B^i_{-1/3}+X^i_{5/3} \\ -i(B^i_{-1/3}-X^i_{5/3}) \\
		T^i_{2/3}+X^i_{2/3} \\ i( T^i_{2/3}-X^i_{2/3}) \\ \sqrt{2}\,\tilde{T^i}_{2/3}
	\end{pmatrix} \, ,
\end{align}
where the subscript indicates the electric charge and here and below the superscript $i,j=2,3$ stand for flavor indices.
In the above formula we classified the composite states according to their quantum numbers
under $SO(4) \simeq SU(2)_L \times SU(2)_R$. In particular the $T_{2/3}$ and $B_{-1/3}$ fields
form an $SU(2)_L$ doublet with $T^3_R$ charge $-1/2$, while $X_{5/3}$ and $X_{2/3}$ belong
to a second doublet with $T^3_R$ charge $+1/2$. The $\widetilde T_{2/3}$ field is instead a singlet.
Notice that, as customary in this class of models, the fermion fields are also charged under an additional unbroken $U(1)_X$ group. The presence
of this extra symmetry is necessary to obtain the correct hypercharges for the fermions. In particular
the hypercharge generator is defined as the combination $Y = T^3_R + X$. This choice fixes the
$U(1)_X$ charge of the fermion fields introduced above to be $X = 2/3$.

In our analysis we will concentrate on the flavor anarchic case. In this scenario, the mixing among the fermion generations arises from the
misalignment between the mass matrix of the composite resonances and that describing the mixing of the elementary fermions with the composite
states. The Lagrangian describing the masses and mixings among the composite states reads
\begin{align}\label{eq:psimass}
	\cL_{\rm composite}
=	M^{ij}_4\left( \bar{\psi}^i_L P_4 \psi^j_R  \right)
	+M^{ij}_1\left( \bar{\psi}^i_L P_1 \psi^j_R  \right) + \textrm{h.c.}\,.
\end{align}
In this formula we used the decomposition of the $SO(5)$ representation $\bf{5}$ in $SO(4)$ multiplets,
\bea
	\bf{5}=\bf{4}+\bf{1}=\bf{(2,2)}+\bf{(1,1)}\,
\eea
and we denoted by $P_{4,1}$ the projectors on the bidoublet and the singlet components respectively. It is important to stress that the Lagrangian in Eq.~\eqref{eq:psimass} is invariant under the full $SO(5)$ global symmetry, as ensured by the CCWZ construction. Of course, the fact that each $SO(4)$ multiplet has a different mass is only a manifestation of the spontaneous symmetry breaking of $SO(5)$ and must not be interpreted as an explicit breaking.

In addition to the mass terms in Eq.~(\ref{eq:psimass}), the effective Lagrangian includes terms that mix the elementary fermions and the
composite states. To write these
we introduce the Goldstone matrix parametrizing the Higgs field:
\begin{align}
	U(H) = \exp\left(i\sqrt{2} h^{\hat a} T^{\hat a}/f \right) \, ,
\end{align}
where $T^{\hat a}$ ($\hat a = 1, \ldots, 4$) denote the genetators of the coset $SO(5)/SO(4)$ and $h^{\hat a}$ are the Higgs components. Notice
that the Goldstone matrix $U(H)$ has special transformation properties under $SO(5)$. On the left it transforms linearly, while on the right it
transforms only with $SO(4)$ elements which correspond to a non-linear realization of $SO(5)$. In the CCWZ formulation the elementary fields
transform linearly under $SO(5)$, while the composite fields transform only under the $SO(4)$ unbroken subgroup. Therefore, to write a mass
(mixing) term that respects the symmetry structure of the theory, we need to multiply the elementary fields by the Goldstone matrix and then couple
them to the composite resonances.

In the ``two site'' model the mixing between the composite and the elementary states is realized as follows
\begin{align} \label{eq:lmixing}
	\cL_{\rm mixing}
=	(\lambda^\dagger_R)^{ij}\, \bar{u}^i_R\, U(H)\, \psi^j + \lambda^{ij}_L\, \bar{q}^i_L\, U(H)\, \psi^j + \textrm{h.c.} \,.
\end{align}
Note that the Lagrangian in Eq.~\eqref{eq:lmixing} formally preserves a bigger symmetry than just $SO(5)$.
In fact it is formally invariant under two independent $SO(5)$ symmetries: one acting only on the elementary fields and
one only on the composite fields (under this symmetry the Goldstone matrix should transform as a bi-fundamental,
see~\cite{Panico:2011pw} for more details). In the most general case in which one only imposes the usual
$SO(5)$ symmetry, one could also add independent mixing terms
for each elementary field to the bidoublet and singlet components of $\psi$.
For instance for the $q_L$ doublet we could write $
\lambda_4\bar{q}_LU(H)P_{4}\psi+\lambda_1\bar{q}_LU(H)P_{1}\psi$, while in Eq.~(\ref{eq:lmixing}) they are set to be equal
\bea
	\left(\lambda_4\right)_{L,R}=\left(\lambda_1\right)_{L,R}=\lambda_{L,R}\,.\label{lambda}
\eea
We adopted the choice in Eq.~\eqref{lambda} because it ensures that the Higgs mass is calculable
in our model~\cite{Panico:2011pw,Matsedonskyi:2012ym}. The most general Lagrangian, instead,
leads to a divergent Higgs mass.

The interactions of the fermions with the $Z$ boson arise from the following terms in the effective
Lagrangian~\cite{Grojean:2013qca}
\begin{align}
	\cL_{Z,\, {\rm int}}
	\label{ltcz}
=	\frac{g}{c_W}\sum_{q=q^{t,c}_L,t(c)_R}\bar{q}\left( T^3_L - Q_q s_W^2\right) \slashed{Z} q
	+\sum_i \bar{\psi}^i \hat{\slashed{e}} \psi^i +
	\sum_{i,j} \zeta_{ij} \bar{\psi}^i \hat{\slashed{d}} \psi^j\,,
\end{align}
where $Q_q$ is the electric charge while $c_W$ and $s_W$ denote the cosine and sine of the weak
mixing angle.
Notice that the elementary fields interact with the $Z$ in the same way as in the SM,
while the interaction of the composite fields arises from the CCWZ covariant derivatives $\hat e$
and $\hat d$. The $\hat e$ and $\hat d$ symbols are defined as~\cite{Contino:2011np}
\begin{align} \label{eq:eddefi}
	&\hat e_\mu=- i\sum_{a}T^{a}\, \mathrm{Tr}\left[U(H)^\dagger D_\mu U(H)\, T^{a}\right] \, , \nonumber\\
	&\hat d_\mu=- i\sum_{\hat a}T^{\hat{a}}\, \mathrm{Tr}\left[U(H)^\dagger D_\mu U(H)\, T^{\hat a}\right]\,,
\end{align}
where $D_\mu$ is the SM covariant derivative, while $T^{a}$ ($a = 1, \ldots, 6$) and $T^{\hat a}$ ($a = 1, \ldots, 4$)
are the unbroken and broken $SO(5)$ generators respectively. For simplicity, we work in
the limit of $\zeta\to0$. The case of non-vanishing $\zeta$ leads to additive flavor violation effects and is worth a further investigation. For $\zeta\sim
\mathcal{O}(1)$ these effects are expected to be at the same order as the discussed effects.

It is interesting to notice that the leading effects of the composite  vector fields $\rho$
are already included in the Lagrangian in Eq.~\eqref{ltcz}. The CCWZ formalism, indeed, ensures that the one in
Eq.~\eqref{ltcz} is the most general Lagrangian consistent with nonlinear linearization of $SO(5)/SO(4)$ breaking.
We can easily check this result explicitly by considering the most general Lagrangian
describing a vector resonance $\rho^a_\mu$ in the adjoint representation of $SO(4)$.
The only terms invariant under the global symmetry are
\bea
	{\cal L}_{\rho}
= 	-\frac{1}{4} \rho_{\mu\nu}^a \rho^{a \mu\nu}+ \kappa f^2 (\hat e^{a}_\mu -g_\rho \rho^{a}_\mu)^2
	+ \kappa_\psi\bar \psi (\hat {\slashed e}^{a} -g_\rho \slashed\rho^{a} )T^{a}\psi\,.
\eea
One can see that the non-kinetic terms depend on the same combination of $\hat e_\mu$ and $\rho_\mu$,
so that integrating out the $\rho$ resonance at zero momentum does not lead to any additional
contribution to the $\hat{\slashed e}$ term of the Eq.~\eqref{ltcz}.
Analogously one can show that the leading effects due to an axial vector resonance
in the coset $SO(5)/SO(4)$ can be encoded in the $\hat d$ symbol terms in Eq.~\eqref{ltcz}.

\subsection{The anarchic flavor structure}

In this subsection we will briefly review the structure of ``flavor anarchic'' composite Higgs models,
which is the basis of our analysis of the $Ztc$ coupling.
The starting point to understand this setup is the partial compositeness
assumption, which links the generation of the SM fermion masses to the mixing of
the elementary states with the composite resonances. To leading order in the elementary-composite mixing terms, the masses of the
SM quarks gives
\bea\label{eq:SM_masses}
m^{i}_{\rm SM} \sim \frac{v}{f} \frac{\lambda^i_L \lambda_R^i}{M_*} ,
\eea
where $M_*$ is the mass scale of the fermionic composite states and $\lambda^i_X$ stands for the $i$th eigenvalue of the matrix $\lambda^{ij}
_X$\,.

The partial compositeness scenario provides an interesting framework
to generate the hierarchical structure of the SM fermion masses and of the
CKM matrix elements. The construction work as follows.
First of all we assume that the strongly interacting sector has an anarchic
flavor structure, that is the mass matrix of the composite states is
generic and all its elements are of the same order, including the off-diagonal ones.
This assumption implies that no flavor symmetry is present in the composite sector, so that the
fermion mass eigenstates are an admixture of the partners of the different generations.
As a second ingredient we require that all the flavor hierarchies of the SM
quark masses and of the CKM elements are generated by the eigenvalues of the elementary composite
mixing matrices.

The requirement of reproducing the correct quark masses and the CKM matrix leads
to the approximate relations between the elementary--composite mixing
parameters $\lambda_{L,R}^i$ presented in the introduction, see Eq.~\eqref{eq:flavor_anarchy}.
This type of ansatz for the elementary composite mixing is usually referred to as ``flavor anarchy''.
As can be seen from Eq.~\eqref{eq:flavor_anarchy}, in this scenario, the mixing of the
light generation quarks to the composite states is small, or, in other words, the first
and second generation quarks are almost elementary.

One of the nice features of the flavor anarchic scenario is the fact that it automatically provides a partial
suppression of the flavor violating NP effects~\cite{Agashe:2006wa}. The reason for this is simply
the fact that flavor changing currents can only be generated through the mixing of the SM states
with the composite sector. Therefore, any effect involving the light quarks is necessarily suppressed
by their small amount of compositeness.

\subsection{The flavor structure of minimal pNGB models}\label{FlavorpNGB}

The flavor structure of composite pNGB models is somewhat different than that of conventional composite models~\cite{Agashe:2004cp} and thus
we are going to consider it in some detail in this part. On the elementary side the flavor group is obviously identical to that of the SM, $SU(3)_Q\times SU(3)_U\times SU(3)_D$ for the quark doublets and up and down singlets respectively. Here, we shall focus on the minimal composite model. In this case we can define in principle the same group structure for the composite sector, assuming that the quarks form fourplets and up and down singlets of $SO(4)$, $SU(3)_{Q^4}\times SU(3)_{U^1}\times SU(3)_{D^1}$ (the superscripts stands for the $SO(4)$ representation).

We can now collect the various flavor violating terms from Eqs.~\eqref{eq:psimass},~\eqref{eq:lmixing} and~\eqref{ltcz} (where as already
mentioned we focus for simplicity on a finite pNGB model~\cite{Panico:2011pw,Matsedonskyi:2012ym}). The vector like composite masses~\eqref{eq:psimass}, $M^{ij}_{4,1}$, transforms as adjoints of the corresponding composite flavor group (here for simplicity we only consider the
terms relevant for up flavor violation). The $\zeta$ term~\eqref{ltcz} transforms as a bi-fundamental of the composite flavor group $SU(3)_{Q^4}\times
SU(3)_{U^1}$\,, however, for simplicity we set it to zero in what follows. Recall that in our construction, the mixing terms~\eqref{eq:lmixing}, $
\lambda_{L,R}$ are non-generic. The choice made in Eq.~\eqref{lambda} to use a mixing with an additional $SO(5)$ symmetry implies that $\lambda_{L,R}$ transform
simultaneously as vector like masses (hence as bi-fundamental of $SU(3)_{Q,U}\times SU(3)_{Q^4,U^1}$), and as Yukawa terms (hence as
bi-fundamental of $SU(3)_{Q,U}\times SU(3)_{U^1, Q^4}$)\,. This implies that they break the $SU(3)_{U^1}\times SU(3)_{Q^4}$ down to a diagonal
vector symmetry $SU(3)_{\psi}$, under which we identify the following spurions, transforming under $SU(3)_Q\times SU(3)_U\times SU(3)_\psi$:
 \bea
 	M^{ij}_{4,1}\in (\mathbf1, \mathbf1,\mathbf8+\mathbf1)\,, \ \ \ \ \ \
	\lambda_L \in (\mathbf 3, \mathbf1,\bar {\mathbf3}) \,, \ \ \ \ \ \
	\lambda_R \in (\mathbf 1, \bar{\mathbf3},\mathbf3)\,.
 \eea

In order to get a first non-trivial usage of the above spurion flavor description let us have a look at the flavor structure of the SM masses within
minimal models. For simplicity, we shall stick to the limit of small mixing, adding higher power in the mixing matrices $\lambda_{L,R}$ is not
expected to change the qualitative features of the result.
Written in matrix form the mass matrices for the fermions can be written as (see {\it e.g.}~\cite{Delaunay:2013pwa}):
\beq
 	M_u^{ij}
	=\left(\begin{array}{cccc}
                0 & \lambda_L \cos^2\frac{\epsilon}{2} &  \lambda_L \sin^2\frac{\epsilon}{2} & -\frac{ \lambda_L }{\sqrt{2}}\sin\epsilon \\
		 \frac{ \lambda_R }{\sqrt{2}}\sin\epsilon & M_4 & 0 & 0 \\
		 -\frac{ \lambda_R }{\sqrt{2}}\sin\epsilon & 0 & M_4 & 0 \\
		  \lambda_R \cos\epsilon & 0 & 0 & M_1
               \end{array}\right)^{ij}\,,\quad\quad \epsilon \equiv \frac{v}{f}\,,
\label{uMassmat}
\eeq
with $M_u$ being mass matrix of the charge $2/3$ states, given in the following basis
\beq
	 \mathcal{L}_{\rm mass}
=	-\left(\bar{u}\ \bar{T}\ \bar{X}_{2/3}\ \bar{\tilde{T}}\ \right)_L^i M^{ij}_u
	 \left(\begin{array}{c}
           u\\
           T\\
           X_{2/3}\\
           \tilde{T}
       \end{array}\right)^j_R   +\mbox{h.c.}\,,
 \eeq
The resulting leading order spurion expression for the SM masses is then
\beq
	\left(m_u^{\rm SM}\right)^{ij}
\simeq \frac{\epsilon}{\sqrt{2}} \lambda_L^{ik} \left[\left(M^{-1}_4\right)^{kl} - \left(M^{-1}_1\right)^{kl}\right] \lambda_R^{lj} \,.\label{massspurion}
\eeq
One can easily check that, for small mixing and anarchic $M_{4,1}$ the above expression recovers the single generation expression given in Eqs.~\eqref{eq:SM_masses} and~\eqref{chirmass}. Furthermore, in the global $SO(5)$ limit, $M_4=M_1$, the SM masses vanish as expected as in that case one can use the approximate global $SO(5)$ symmetry to rotate $\psi$ in such away that no dependence on $H$ is present either in the mixing terms in Eq.~\eqref{eq:lmixing} or in $M_{4,1}$. From Eq.~\eqref{massspurion} we learn that, similarly to the non-pNGB Higgs case, the hierarchies in masses are controlled by the hierarchies in the eigenvalues of $\lambda_{L,R}$~\cite{Grossman:1999ra,Huber:2000ie,Gherghetta:2000qt}. However, the source of flavor violation (beyond the SM) in this theory is due to the misalignment, in flavor space, between $\lambda_{L}^\dagger \lambda_{L},\, \lambda_{R} \lambda_{R}^\dagger$ and $M_{4,1}$ (to be compared with the non-pNGB case~\cite{Agashe:2004cp}).

Let us count the number of physical mixing angles in the two generation case. This will become handy when we discuss $t\to c$ transition and
when we examine the potential linkage between flavor violation and anarchic naturalness. To further simplifies the discussion, and as we are not
interested in CP violation let us switch off all the complex parameters. In addition, we shall only focus on the up-type flavor sector that is relevant to
top flavor violation.
When switching off the masses and the linear mixing terms, the Lagrangian admits the following large $SO(2)^6$ symmetry group
$SO(2)_{Q,U}\times SO(2)_{Q^4,U^1}^{L,R}$.

Generically, the Lagrangian consist of six independent $2\times 2$ real matrices $M_{4,1}$ and $\left(\lambda_{4,1}\right)_{L,R}$\,. As explained to
ensure the finiteness of the potential we shall assume that the $\lambda$'s respect the $SO(5)$ symmetry and respect the ``pseudo-covariant"
relation (in flavor space) given in Eq.~\eqref{lambda} $\left(\lambda_4\right)_{L,R}=\left(\lambda_1\right)_{L,R}$\footnote{We note that the general
condition for finite potential is ${\rm Tr}[(\lambda_1)_{L,R} (\lambda_1)^\dagger_{L,R}]={\rm Tr}[(\lambda_4)_{L,R} (\lambda_4)^\dagger_{L,R}]$.}.
We call this pseudo-covariant as it holds in any basis when considering the elementary flavor space but not on the composite side that does not
respect the $SO(5)$ symmetry, namely the $M_4$ and $M_1$ are independent matrices.
This is an interesting manifestation of the fact that the flavor group does not commute with the $SO(5)$ symmetry.

The above covariant relation implies that there is a single elementary (unphysical) mixing angle for $\lambda_L$ and $\lambda_R$ (that can be removed via the
$SO(2)_{Q,U}$ transformation) and that the two eigenvalues of $\left(\lambda_4\right)_{L,R}$ and $\left(\lambda_1\right)_{L,R}$ (that are basis
independent) are the same. This results in 18 physical parameters (24 minus 6). The $SO(2)^6$ symmetry transformation allows to remove 6
unphysical flavor parameters leaving total of 12 independent flavor parameters. Out of which we can identify the 8 eigenvalues of $\lambda_{L,R}$
and $M_{1,4}$ and two pair of composite physical mixing angles, $\theta_{L,R}^{1,4}$. These correspond to the misalignment between $
\lambda_{L,R}$ and $M_{1,4}$ respectively (omitting transpose signs for simplicity).

Given the above description, in the basis where $\lambda_{L,R}$ are diagonal we use the following parametrerization
\begin{align} \label{eq:defMLam}
	M_{1,4} = O^{1,4}_L \hat{M}_{1,4} (O^{1,4}_R)^\dagger  \,, \qquad
	\lambda_{L,R} = \hat{\lambda}_{L,R}\, ,
\end{align}
where $\hat{X}$ are the diagonal matrices
\begin{align} \label{eq:defMx}
&	\hat{M}_{X} = {\rm diag}\left[ M_{X_c}\, , \ M_{X_t}\right] =M^*_{X}\times {\rm diag}\left[ 1+\Delta_X\, , 1 \right] \, , \quad
	\hat{\lambda}_Y = {\rm diag}\left[ \lambda^c_Y, \lambda^t_Y \right] \, ,
\end{align}
and $O^X_Y$ are the relative rotations
\begin{align}
&	O^{X}_Y =
	\begin{pmatrix} \cos(\theta^X_Y) & \sin(\theta^X_Y) \\
	-\sin(\theta^X_Y) & \cos(\theta^X_Y) \end{pmatrix} \, ,
\end{align}
with $X=1,4$ and $Y=L,R$.

\subsection{The Higgs potential and top compositeness}  \label{sec:higgspot}
\subsubsection{Single generation case}
Before proceeding further we will review the properties of the Higgs potential in the MCHM$_5$
framework. For this short discussion we will follow the lines of~\cite{Matsedonskyi:2012ym,Panico:2012uw} that is analysing the potential focusing
on the contribution from the top sector.
We shall include the contributions from the second generation in the following subsection.
As we will see the analysis of the Higgs potential is essential to determine the
most natural part of the parameter space of the model
and leads to important consequences on the flavor phenomenology.
The main contributions to the composite Higgs potential are generated at the radiative level
through loops of the composite top partners. The mixing of the elementary top components, indeed,
induces the largest breaking of the $SO(5)$ global symmetry which protects the Higgs dynamics.

Due to the Goldstone nature of the Higgs, the effective potential can be always expressed
in terms of trigonometric functions of the ratio $h/f$. In the following we will be interested
in the configurations with $(h/f)^2 \ll 1$, which are preferred by EW precision data~\cite{Grojean:2013qca,Ciuchini:2013pca} and by the current bounds on
the Higgs couplings~\cite{CMS:2014,ATLAS:2014}. In this case it is convenient to use an expansion of the potential in powers of $\sin(h/f)$. The Higgs potential can
therefore be rewritten in the form
\begin{equation}\label{eq:potential}
	V(h)=\alpha \sin^2 \left(\frac{h}{f}\right)+\beta \sin^4 \left( \frac{h}{f} \right) \,.
\end{equation}
The position of the minimum of the potential is determined by the ratio of the $\alpha$ and $\beta$ coefficients:
\begin{align}
\xi\equiv \sin^2\left(  \frac{\langle h \rangle }{f} \right)=-\frac{\alpha}{2\beta}\,.  \label{xi}
\end{align}

We can now use the above results to get an estimate of the fine tuning of the model.
The main source of tuning is related to the requirement $\xi \ll 1$. For generic
values of $\alpha$ and $\beta$, indeed, the minimum of the potential in Eq.~(\ref{eq:potential})
is naturally found at $\xi \sim 1$. Some amount of cancellation in the $\alpha$ coefficient
is thus needed to get a phenomenologically viable configuration.

To get a more quantitative estimate of the tuning, first of all we need to get an estimate of the size of the $\alpha$ and $\beta$ coefficients. The fine-tuning can then be estimated as the ratio between the actual value of the $\alpha$ needed to obtain the correct $\xi$, given in Eq.~\eqref{xi}, and the typical size of each of the term contributing to $\alpha$. The $\alpha$ and $\beta$ coefficients are induced by the
breaking in the top sector. The $\alpha$ coefficient is of the order of
 \begin{equation}\label{eq:alpha_beta_estimates}
	\alpha^t_{L,R} \sim \frac{N_c}{16 \pi^2} (\lambda^t_{L,R})^2 M_*^2\,,
\end{equation}
whereas $\beta$ is typically of the order of
 \begin{equation}\label{eq:alpha_beta_estimates}
	\beta \sim \frac{N_c}{16 \pi^2} (\lambda_L^t \lambda_R^t)^2\,,
\end{equation}
where $N_c = 3$ is the number of QCD colors.
With these definitions we get
\bea
	\hbox{FT}^{-1}\sim \frac{-2\beta\,\xi}{{\alpha^t_{L,R}}}\sim \frac{{\rm min}\left[(\lambda_R^t)^2, (\lambda_L^t)^2\right] v^2}{M_*^2 f^2}\,.
\eea
As we discussed before, see Eq.~(\ref{eq:SM_masses}), to leading order in the
mixings the top mass is $m_t\sim \lambda^t_L\lambda^t_R v / M_* f \,.$
By combining the above relations we find our estimate for the amount of fine tuning:
\bea  \label{FTest}
	\hbox{FT}^{-1}\sim  \frac{m_t^2}{{\rm max}\left[\lambda_{L,R}^t\right]^2} \, .
\eea
To minimize the amount of tuning one typically requires $\lambda_{R}^t\sim \lambda_{L}^t$~\cite{Panico:2011pw}.
Thus, the relation for the fine tuning can be rewritten as
\bea \label{FTest1}
	\hbox{FT}^{-1}\sim  y_t \, \frac{v^2}{M_* f}\sim \frac{y_t}{g_*}\,\frac{v^2}{f^2}\,,
\eea
where $y_t\sim1$ is the top Yukawa and we use $M_*\sim g_*f$.

From Eq.~(\ref{FTest1}) we can see that a small value of $g_*$, or in other words
a small mass scale for the fermionic resonances $M_*\equiv g_* f$, implies a smaller amount
of tuning and improves the naturalness of the model.
An interesting consequence of the presence of light fermionic partners is the fact that
both $t_L$ and $t_R$ necessarily have a large amount of compositeness as follows.
In models with a pNGB Higgs, with linear mixing, we expect that the SM-chiral fermion masses are given by trigonometric function of the Higgs and
thus bounded from above.
For example in minimal models the top mass is given by (see {\it e.g.}~\cite{Delaunay:2013pwa})
\beq
	m_t\approx g_* \,v\,  s^t_L s^t_R\label{chirmass}\,,
\eeq
with $s^t_{L,R}= \lambda_{L,R}^t/\sqrt{M_{4,1}^2+\left(\lambda_{L,R}^t\right)^2}\,,$ where to leading order in the mixings and assuming no
hierarchies in the spectrum,  this expression coincide with the one given in Eqs.~\eqref{eq:SM_masses} and~\eqref{massspurion}. From Eq.~\eqref{chirmass} we first learn that even for large values of $g_*$ both top chiralities are required to be fairly composite.
Furthermore, as already discussed to minimize the tuning $g_*$ is required to be small and  $s^t_L \sim s^t_R$. Therefore both the LH and RH
components of the tops are required to have large, order one, mixing with the composite sector.
This, rather generic conclusion, leads to interesting and important consequences for top flavor violation as we discuss in the following section.

\subsubsection{Higgs potential, the anarchic  case}

In the above, we discussed the properties of the Higgs potential and the amount of fine tuning present in the MCHM$_5$ set-up. In our analysis,
however, we completely ignored the effects of the charm partners. We want now to understand if these fermionic states play a role in the fine tuning
determination.

In the anarchic model all the partners of the light quarks mix strongly
with the top quark and thus their contribution to the Higgs potential is expected to
be comparable. To understand the impact of the charm partners it is useful to analyze
their contribution to the $\alpha$ coefficient in the expansion of the potential
in Eq.~(\ref{eq:potential}). In fact, this coefficient can be used to derive an estimate of the amount of tuning.
In the two-site model the $\alpha$ coefficient is generically logarithmically divergent . This divergence
is regulated by the higher level of resonances that are not included in the two-site
description.\footnote{
A fully calculable Higgs potental can be obtained in a three-site set-up in which one additional layer of composite states is included ~\cite{Panico:2011pw}
(a similar set-up is interpreted as a ``two-site'' in \cite{DeCurtis:2011yx}), although there are some special points in the parameter space of the two-site model with finite potential as well \cite{Carena:2014ria}.
}
In a first approximation, however, the logarithmically divergent piece can be used as an estimate of
the overall size of the $\alpha$ coefficient. An explicit computation gives
\begin{align}  \label{alphamh}
	\alpha
	\approx&\frac{N_c}{16 \pi^2}{\rm Tr}\left[
	\lambda_L^\dagger\left( M_1M^\dagger_1- M_4M^\dagger_4 \right)\lambda_L
	-2\lambda_R \left( M^\dagger_1M_1-M^\dagger_4M_4 \right)\lambda_R^\dagger \right]  \log\Lambda^2/M^2_*  \, ,
\end{align}
where $\Lambda$ is the cut off scale. For the case of only second and third generations and in the limit $\lambda^t_{L,R}\gg\lambda^c_{L,R}$,
$\alpha$ is estimated to be
\begin{align} \label{eq:alphamh2}
	\alpha\approx&   \frac{N_c}{16\pi^2}\left( \alpha_t +\alpha_{\rm mixing} \right)\log\Lambda^2/M^2_* \, ,
\end{align}
with
\begin{align}
	\alpha_t = &\left[  (\lambda^t_L)^2 -2(\lambda^t_R)^2\right]  \left( M^2_{1_t} - M^2_{4_t} \right) \, , \nonumber\\
	\alpha_{\rm mixing} =
	&(\lambda^t_L)^2\left[  \left( M^2_{1_c} - M^2_{1_t} \right)\sin^2(\theta_L^1)
	- \left( M^2_{4_c} - M^2_{4_t} \right)\sin^2(\theta_L^4)\right] \nonumber\\
&-2	(\lambda^t_R)^2\left[   \left( M^2_{1_c} - M^2_{1_t} \right)\sin^2(\theta_R^1)
	- \left( M^2_{4_c} - M^2_{4_t} \right)\sin^2(\theta_R^4)   \right]  \, ,\nonumber
\end{align}
where the fine-tuning due to the top partner is proportional to FT$_{t}\propto \alpha_t$ and FT$_{\rm mixing}\propto\alpha_{\rm mixing}$ is due to top-charm mixing with
\beq
{\rm FT} \approx {\rm FT}_{t}+{\rm FT}_{\rm mixing}\,.
\eeq
Notice that the leading contributions to $\beta$, the quadratic piece of the potential given in Eq.~\eqref{eq:alpha_beta_estimates}, are
proportional to the forth power of
the mixing, with no dependence on the resonance mass parameters. As $\beta$ has to be flavor single the only dependence of $\beta$ in terms of
the mixing parameters has to be proportional to ${\rm Tr} \left[\lambda_{L,R} \, \lambda_{L,R}^\dagger\right]$. Therefore, to leading order it is
independent of the additional top-charm mixing parameters.

From Eqs.~\eqref{alphamh}--\eqref{eq:alphamh2} it is clear that, when the second level of resonances is much heavier than the first one,
$M^2_{X_c} \gg M^2_{X_t}$, the size of the $\alpha$ coefficient becomes larger, especially if the mixing between the second and third generations
is sizable as expected in anarchic models, $\sin(\theta_{L,R}^{1,4}) \sim 1$. As mentioned above, a larger natural value for $\alpha$ translates in
an increase in the fine tuning. As we further discuss below, this forms an interesting conceptual (though admittedly weak in general) linkage between the physics of flavor and the physics of
naturalness.

This result can also be understood in an equivalent way as follows. Let us consider a minimal
effective description in which all the composite resonances have been integrated out and only the
elementary states and the $\sigma$-model describing the Higgs are retained. In this case
the $\alpha$ coefficient in the Higgs potential is quadratically divergent.
In a more complete description including the composite dynamics, this divergence is
regulated by the composite resonances thorugh a collective breaking mechanism
and only a logarithmic divergence is left as we saw in the two-site model~\cite{Panico:2011pw}.
It is thus clear that the natural size of the $\alpha$ coefficient
is set by the mass of the resonances that
cut-off the quadratic divergence coming from the top loop (see Eq.~(\ref{eq:alpha_beta_estimates}))\footnote{Instead of introducing the new resonances one can use the holographic approach and parametrize all the effects of the composite fields by modified correlators of the two point functions \cite{Contino:2006qr}.  In this case the quadratic divergences are cut at the scale when momentum dependent corrections  to the two point function correlators become important, which  happens at the top partner mass scale.  }.
Of course in a two-site model with only top partners the first level of resonances is quadratic
enough to regulate the  divergence. On the other hand, if we add the charm partners and we assume that
they have a sizable mixing with the top, part of the quadratic divergence will not be regulated
any more by the top partners but instead by the new states. If a mass gap exists between
the two sets of states, then part of the quadratic divergence will be regulated at a higher cut-off
and the value of $\alpha$ will necessarily increase. This mechanism is clearly recognizable
in the explicit expression in Eq.~(\ref{eq:alphamh2}). The terms proportional to $M^2_{X_c} - M^2_{X_t}$
correspond to the contribution of the additional partners as can be deduced by the fact that they
are weighted by the top--charm mixing angles $\sin(\theta_{L,R}^{1,4})$.

We finally mention that due to the smallness of the linear-mixing terms into the first and second generations we expect that all the three resonances after produced will decay to third generation SM field.
It implies that not only that the fine tuning of anarchic models get worsen once all the three generations of partners are included but also the bounds on the top partners are effectively stronger than those that are commonly extracted within the single generation case (due to the large production cross section).

\section{Top flavor violation} \label{sec:MCHM5}

\subsection{$t\to cZ$ transition}

Now we proceed to the calculation of the $Ztc$ interactions in the MCHM$_5$. We need to match this model's specific parameters to the generic
ones that we have already discussed in Eqs.~\eqref{eq:dim6}--\eqref{generic}. We can easily estimate the importance of the three relevant
operators of Eq.~\eqref{eq:dim6} by means of spurion analysis. This can be done by promoting the elementary--composite mixing parameters
defined in Eq.~\eqref{eq:lmixing} to spurions formally transforming in the fundamental representation of $SO(5)$ (see {\it e.g.}~\cite{Panico:2011pw} for more datails). The physical values of the spurions can be rotated in $SO(5)$ space to take the following background value
\begin{align}
&	\lambda_L^{ij}  \ra \frac{\lambda^{ij}_L}{\sqrt{2}}\, (0,0,1,i,0)^T \,  , \qquad
	\lambda_R^{ij} \ra \lambda_R^{ij}\, (0,0,0,0,1)^T \, ,
\end{align}
where we took into account only the charge $2/3$ components of the elementary doublets
$q_L^{i}$ that are the only ones needed for our analysis.

We can now construct the structures that contribute to the $Ztc$ interactions and generate the operators in Eq.~\eqref{eq:dim6}. First of all we
consider the coupling involving the left-handed fermions. To leading order in the spurions $\lambda_{L,R}$ the $Z$ coupling matrix to the left-handed
matrix can be written as
\begin{align} \label{eq:gZL}
	\left(g^{\rm SM}_{Z,L}\right)^{ij}
=&	-\frac{g}{2c_W}\left[ \frac{4s^2_W-3}{3}\delta^{ij}
	+\frac{\epsilon^2}{2}\lambda^{ik}_L
	\left( \left(M^\dagger_{1}M_{1}\right)^{-1} + \left(M^\dagger_{4}M_{4}\right)^{-1}  \right)^{kl}
	(\lambda_L^\dagger)^{lj}\right] \, .
\end{align}
In order to estimate the resulting flavor violation we need to move to the mass basis for the LH SM fields.
The mass basis is defined via the basis in which the spurion $A_{LL}\equiv m_u^{\rm SM}  \,m_u^{\dagger \rm SM}$, defined in Eq.~\eqref{massspurion}, is diagonal. Within the anarchy paradigm both $A_{LL}$ and $g^{\rm SM}_{Z,L}$ are hierarchical and approximately aligned as dictated by ``RS-GIM"~\cite{Agashe:2004cp}. They are, however, slightly misaligned in flavor space, with the mixing angles that control the flavor violation of order of the ration of the eigenvalues of the $\lambda_L$ as in Eq.~\eqref{eq:flavor_anarchy}. The flavor triviality limit, where no flavor violation occurs, is achieved by the alignment of both $M_{1,4}$ with $\lambda_{L,R}$ in the flavor space, i.e. the limit $\theta^{1,4}_{L,R}\to0$.

\begin{figure}
\begin{center}
\includegraphics[height=.175\textwidth]{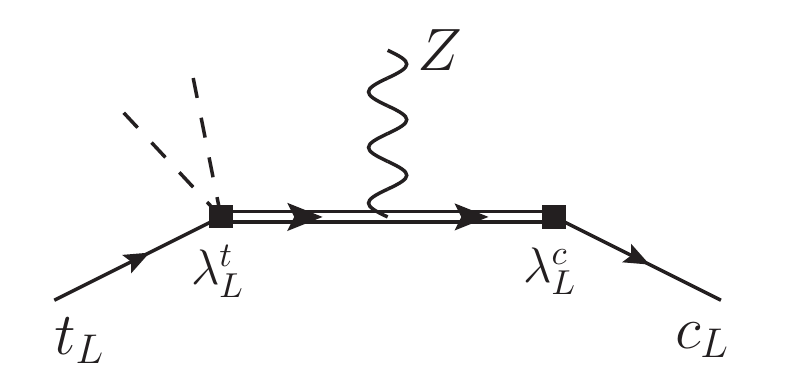}
\hspace{1em}
\includegraphics[height=.175\textwidth]{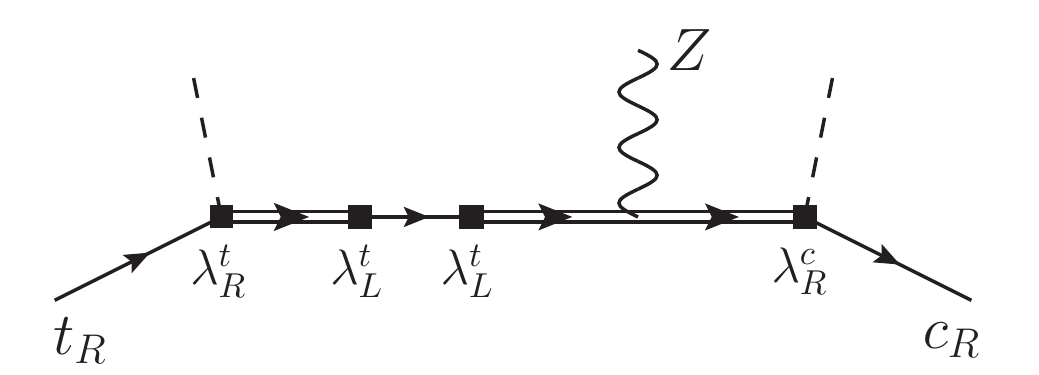}
\end{center}
\caption{Schematic structure of the diagrams contributing to the flavor violating
$Z$ couplings with the top and the charm quarks. The single lines denote the elementary
fields while the double lines correspond to the composite states. Each dashed line
denotes one insertion of the Higgs VEV.}\label{tczFD}
\end{figure}

Following the above discussion we can estimate the contribution to the $g_{tc,L}$  couplings, see also Appendix~\ref{app:diagM}:
\begin{align}\label{eq:g_spurL}
	 g_{tc,L}
& \sim \frac{g}{2 c_W} \frac{v^2}{2f^2} \frac{\lambda_L^t}{M_*}\frac{\lambda_L^c}{M_*}
\sim	 \frac{g}{2\sqrt{2} c_W} \frac{v}{f}\frac{m_t }{M_*}V_{cb}
\sim 8.1 \times 10^{-4} \l(\frac{700\, {\rm GeV} }{f}\r)\l(\frac{700\,{\rm GeV}}{M_*}\r)\,,
\end{align}
where in the last step we used the ``minimal  tuning'' assumptions from the Eqs.~\eqref{FTest}
and~\eqref{FTest1} and the flavor anarchy relations in Eq.~\eqref{eq:flavor_anarchy}.
The schematic structure of the Feynman diagram that gives rise to the above coupling
is shown on the left of Fig.~\ref{tczFD}. We see that the expression in Eq.~\eqref{eq:g_spurL} nicely fit with the generic expressions given in
Eqs.~\eqref{basis} and~\eqref{generic}.

Let us now analyze the flavor changing coupling involving the right-handed quarks. If we naively substitute the $\lambda_L$ spurions with the $\lambda_R$ ones and derive the analogous of Eq.~(\ref{eq:gZL}), we find out that the non-universal piece vanishes. The reason for the cancellation is the fact that the custodial $P_{LR}$ symmetry, that protects the $Z$ coupling to the $b_L$ quark, also affects the $t_R$ and the $c_R$ fields~\cite{Agashe:2006at}. The $t_R$ and $c_R$ components, indeed, belong to the singlet representation of $SO(4)$, thus they are trivially invariant under the exchange of the $SU(2)_L$ and $SU(2)_R$ subgroups of $SO(4)$ (this invariance is dubbed $P_C$ parity \cite{Agashe:2006at}). As a consequence of the $P_{C}$ symmetry the corrections to the $t_R$ and $c_R$ couplings to the $Z$ boson can only be generated through the insertion of the $P_{C}$ breaking couplings, namely $\lambda_L^t$ (or $\lambda_L^c$, which is however much smaller and leads to a sub-leading correction). Thus, the leading order contribution (in term of the suporion) to the $Z$ couplings matrix to the right-handed fermion is
\begin{align}\label{eq:gZR}
	\left(g^{\rm SM}_{Z,R}\right)^{ij}
=	-\frac{g}{2c_W}\left[ \frac{4s^2_W}{3}\delta^{ij}
	+\frac{\epsilon}{\sqrt{2}} \left(
	m^{\rm SM\dagger}_u\lambda_L \left( M_{4} M_{4}^\dagger M_{4}\right)^{-1}\lambda_R +h.c. \right)^{ij}\right] \, ,
\end{align}
where $m_u^{\rm SM}$ should be taken as the spurion that appears on the right hand side of Eq.~\eqref{massspurion}.

The corresponding estimate of the $g_{tc,R}$ coupling is
\begin{align}\label{eq:g_spurR}
	g_{tc,R}
&\sim \frac{g}{2c_W}\frac{v^2}{f^2}\frac{\lambda_R^t}{M_*}\frac{\lambda_R^c}{M_*} \left(\frac{\lambda^{t}_{L}}{M_*} \right)^2
\sim  \frac{g}{2 c_W} \frac{1}{M_*^2}\l(\frac{m_c m_t}{V_{cb}}\r)
\sim 1.5 \times 10^{-3} \l(\frac{700\,{\rm GeV} }{M_*}\r)^2\,,
\end{align}
see also Appendix~\ref{app:diagM}, as for the left-handed coupling an additional contribution comes from operators containing the
bidoublet mass matrix. Note that the additional suppression given by the $(\lambda_L^t/M_*)^2$ factor is generic for all models that use the
custodial symmetry to protect $Z \bar b_L b_L $ (and $Z \bar s_L s_L$) coupling. In all these models, indeed, the $t_R$ and $c_R$ fields must be
in custodially protected representations~\cite{Agashe:2006at}.

Before concluding this section it is useful to comment on the phenomenological implications of the
custodial protection for the right-handed coupling. With respect to a model without custodial
protection, the $g_{tc,R}$ coupling is suppressed by two powers of the left-handed top compositeness
angle $s_L^t \sim \lambda_L^t/M_*$. As discussed at the end of subsection~\ref{sec:higgspot}, the $t_L$ compositeness is tightly related to the mass scale of the composite resonances $m_\psi = g_* f$ and must satisfy the lower bound $s_L^t \gtrsim y_t/g_*$.
This means that in natural scanerios, that require light resonances ($g_* \lesssim 2$),
the additional factor in $g_{tc,R}$ does not lead to any signiuficant suppression. The reduction
of the right-handed flavor-changing effects is only effective when the composite resonances are
heavy. An explicit confirmation of this can be found in the context of the extra-dimensional
composite Higgs realizations. In that case the mass scale of the fermionic resonances is connected
to the one of the gauge resonances, which are constrained to be rather heavy from the EW data.
This of course implies that a significant suppression of the $g_{tc,R}$ coupling is expected
in custodially-protected models. 

From the above results we can derive the following estimate for the branching fraction ${\rm BR}(t \rightarrow cZ)$
\begin{align}\label{eq:BR_tcZ}
	\BR(t\to c Z)
&	\approx 3.5 \left(\frac{g}{2 c_W}\right)^2\frac{(m_t v)^2}{M_*^4}  \l(\left(\frac{m_c  }{v V_{cb}}\right)^2\,,\,\frac{V_{cb}^2}{2}\r)\nonumber\\
&	\sim \left(0.8\,,\, 0.2\right) \times 10^{-5} \,\l(\frac{700\,\rm GeV }{M_*}\r)^4\, .
\end{align}
The estimate in Eq.~(\ref{eq:BR_tcZ}) shows that the natural size of the branching fraction for
the $t \rightarrow cZ$ decay in the presence of light composite resonances is not far from the
current experimental bounds. The present searches indeed set an upper bound $\BR(t\to cZ)<5\times10^{-4}$ at 95\%\,CL~\cite{Chatrchyan:2013nwa}. Although currently not probed, branching ratios of order $10^{-5}$ will be tested at the LHC in the $14\,$TeV run.

 \subsection{Flavor violation vs. fine tuning}

Next, we are aiming at understanding the correlation between amount of fine-tuning and flavor violation. The main insights can be obtained by considering small perturbations of the theory near its flavor-trivial limit, namely when the mixing angles are small and the mass splittings are small. Consequently, we can expand the relevant observables in term of small mixing angles, $
\theta^{1,4}_{L,R}\ll1$, and mass differences over the universal part, $\Delta_{1,4}\ll1$.

From Eq.~\eqref{eq:alphamh2} we see that additional contribution to the fine-tuning, besides the one from the
top-partners, $\alpha_t$, requires both non-degeneracy and non vanishing mixing angles, $\Delta_{1,4}\ne0$ and $\theta^{1,4}_{L,R}\ne0$ respectively. However,
below we show that flavor violation requires only non-vanishing mixing angles. Therefore, we
conclude that $t\to cZ$ may exist without paying the price of additional tuning. Nevertheless we will also identify a well defined limit, when the misalignment in flavor space is vectorial, i.e.
$\theta^{1,4}_L=\theta^{1,4}_R$, in which the flavor violation is correlated with the fine-tuning. For convenience, let us define the vectorial and axial
misalignment directions
\begin{align}
	\theta^X_V=\theta^X_R+\theta^X_L \, , \qquad
	\theta^X_A=\theta^X_R-\theta^X_L \, .
\end{align}

More insight can be obtained by looking at the flavor symmetry breaking pattern of the model. Switching on the universal part of $M_{1,4}$,
$M_{1,4}^*$  in Eq.~\ref{eq:defMx}, breaks
$SU(3)_{U^{1,4}}\times SU(3)_{Q^{1,4}}$ to the vector group. Thus, the orthogonal rotation, namely the axial one, leads to flavor violating
interaction. The second stage of the flavor symmetry breaking is achieved by $\Delta_{1,4}\ne0$. This breaking will leave us with only $U(1)$'s
symmetries and allows for additional contributions to flavor violation. Since only the second breaking involves an additional scale, $\Delta_{1,4}$, it
is correlated with the fine-tuning.

The limit of degenerate $M_{1,4}$, $\Delta_{1,4}\to0$, can be understood as follows. The rotation angles between the basis where $\lambda_{L,R}
$ are diagonal to the mass basis of the SM up-type quarks can be deduced from Eq.~\eqref{massspurion}. These are proportional to $\theta^{1,4}
_{A}\times \lambda^c_{L,R}/\lambda^t_{L,R}$.\footnote{Notice that in the limit $\Delta_{1,4} \rightarrow 0$ the $\theta_V^{1,4}$ angles
are non-physical and drop off in the frmion mass matrices in Eq.~\ref{eq:defMLam}.}
Thus, the leading contribution to the $t_L \to c_L Z$ transition is given by  $(g^{\rm SM}_{Z,L})_{2,2}$
(of Eq.~\eqref{eq:gZL}) times the left rotation angle. The resulting off diagonal $Z$ coupling is
\begin{align} \label{eq:gLtheta}
	g_{tc,L} \approx
&  	-\frac{g\epsilon^2}{4c_W} \frac{(M^*_4)^2+(M^*_1)^2}{M^*_1-M^*_4}  \frac{\lambda^t_L\lambda^c_{L}}{M^*_1M^*_4}
	\left( \frac{\theta^1_A}{M^*_1} - \frac{\theta^4_A}{M^*_4} \right)\, .
\end{align}
The $t_R \to c_R Z$ transition is a bit more complicated, as the fermions mass matrix is also involved, see Eq.~\eqref{eq:gZR}. Since $m^{\rm SM}
_u$ is very hierarchical effectively, the leading contribution to $g_{tc,R}$ is proportional to  $\left[ \mathcal{O}_L \lambda_L O^4_L (O^4_L)^\dagger
\lambda_R \mathcal{O}_R^\dagger\right]_{2,1} \times m_t$, where $\mathcal{O}_{L,R}$ are the rotations to the mass basis. Therefore, the resulting
RH transition is
\begin{align} \label{eq:gRtheta}
	g_{tc,R} \approx
	-\frac{g\epsilon^2}{4c_W}\frac{(\lambda^t_L)^2\lambda^c_R\lambda^t_R}{(M^*_4)^3M^*_1}
	\left(  \theta^4_A - \theta^1_A  \right) \, .
\end{align}
From Eqs.~\eqref{eq:gLtheta}--\eqref{eq:gRtheta} we conclude that $t\to cZ$ is proportional to the axial mis-alignment angles, $\theta^{1,4}_{A}$,
and can occur without additional contributions to the fine-tuning.

The flavor violating contributions which are proportional to $\Delta_{1,4}\,$ are
\begin{align}
	\delta g_{tc,L} \approx&
  	-\frac{g\epsilon^2}{4c_W} \frac{(M^*_4)^2+(M^*_1)^2}{M^*_1-M^*_4}  \frac{\lambda^t_L\lambda^c_{L}}{M^*_1M^*_4}
	 \Bigg[
	\left( \frac{M^*_1-M^*_4}{(M^*_4)^2+(M^*_1)^2}M^*_4(\theta^1_V+\theta^1_A)+\frac{1}{2}(\theta^1_V-\theta^1_A) \right)\frac{\Delta_1}
{M^*_1}\nonumber\\
	&\qquad\qquad\qquad\qquad+\left( \frac{M^*_1-M^*_4}{(M^*_4)^2+(M^*_1)^2}M^*_1(\theta^4_V+\theta^4_A)-\frac{1}{2}(\theta^4_V-
\theta^4_A) \right)\frac{\Delta_4}{M^*_4}
	\Bigg]\, , \\
	\delta g_{tc,R} \approx&
	-\frac{g \epsilon^2}{4c_W}\frac{(\lambda^t_L)^2\lambda^c_R\lambda^t_R}{2(M^*_4)^3M^*_1}
	\left[ \left( 2\frac{M^*_1}{M^*_4}  - 3 \right) (\theta^4_V+\theta^4_A) \Delta_4+ (\theta^1_V+\theta^1_A) \Delta_1\right]\, ,
\end{align}
these are additional contributions to Eqs.~\eqref{eq:gLtheta}--\eqref{eq:gRtheta}. We see that at the vectorial limit, $\theta^X_A\to0$, these are the
only contributions to $t\to cZ$. These are correlated with the fine-tuning estimation of the model. At the small angles and mass differences limit Eq.~\eqref{eq:alphamh2} can be written as
\begin{align}
	\alpha_{\rm mixing}\approx
&	 \frac{1}{2}\left[ (\lambda^t_L)^2 -2(\lambda^t_R)^2 \right]\left[ (M^*_1)^2 \Delta_1(\theta_V^1)^2 - (M^*_4)^2 \Delta_4(\theta_V^4)^2 \right] \, ,
\end{align}
 where we set $\theta^X_A =0$.

\begin{figure}
\begin{center}
\includegraphics[height=.35\textwidth]{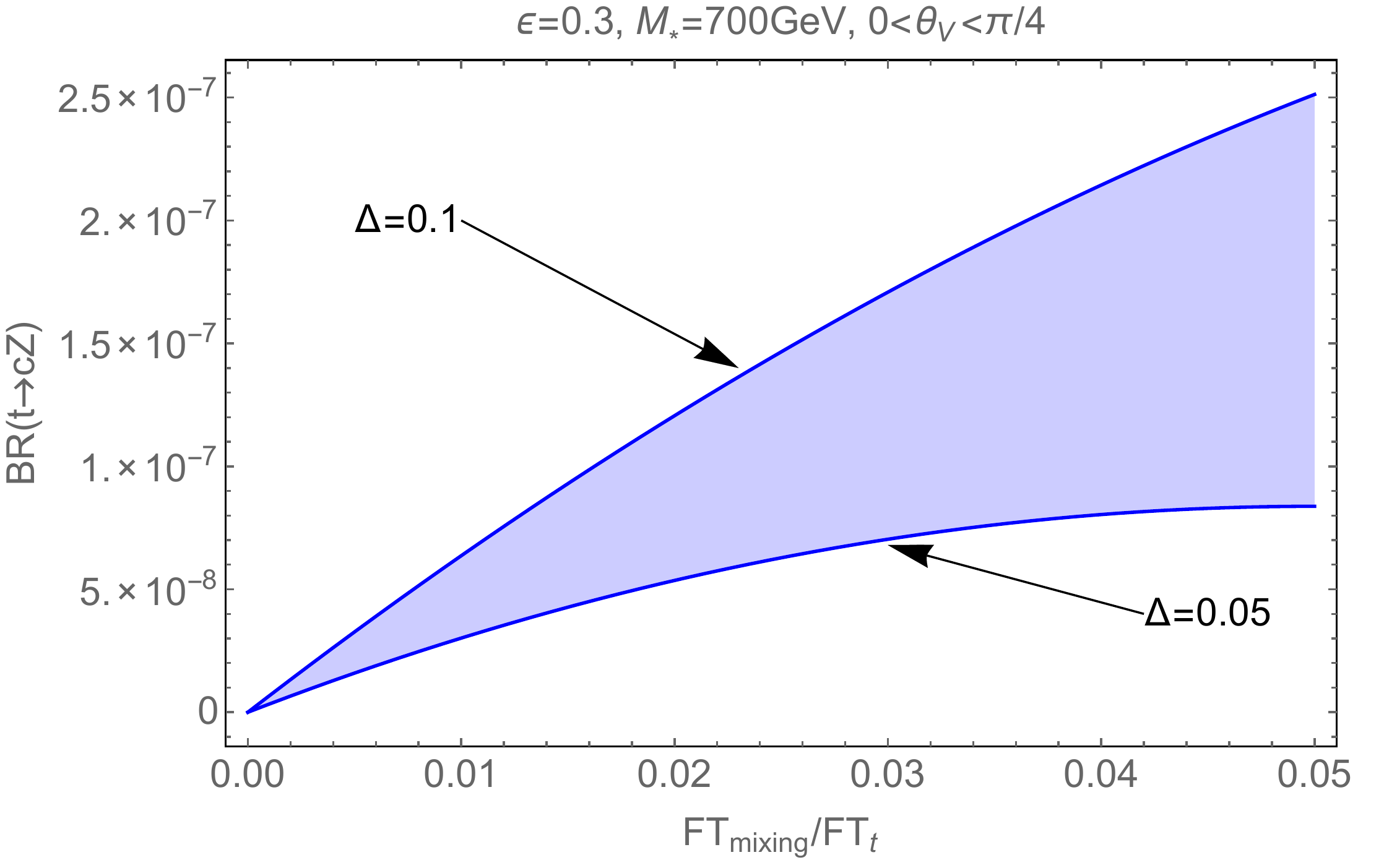}
\end{center}
\caption{The correlation between ${\rm BR}(t\to cZ)$ and the additional fine-tuning of the model ${\rm FT}_{\rm mixing}/{\rm FT}_t$.}
\label{fig:t2czvsFT2}
\end{figure}

To get a bit more insights on the correlation between $t\to cZ$ and tuning price, let us analyze the following simplified case. We consider that case
where $M^*_1=2M^*_4=2M_*$, $\Delta_4=\Delta_1=\Delta$ and switching on only a finite vector like mixing angle with $\theta^4_V=\theta^1_V=\theta_V$. The additional contrition to the  fine-tuning (not from the top-partner) can be estimated as
\begin{align}
	\alpha_{\rm mixing} \approx 6\left[ (\lambda^t_L)^2 -2(\lambda^t_R)^2 \right]M_*^2 \Delta \sin^2(\theta_V/2)
	=  2\alpha_t \Delta \sin^2(\theta_V/2) \, .
\end{align}
 and the $tcZ$ couplings are
\begin{align}
	g_{tc,L} = -\frac{5g\epsilon^2}{32c_W} \frac{\lambda^c_L\lambda^t_L}{M_*^2} \Delta \sin(\theta_V) \, , \qquad
	g_{tc,R} = -\frac{g\epsilon^2}{8c_W}\frac{(\lambda^t_L)^2\lambda^c_R\lambda^t_R}{M_*^4} \Delta \sin(\theta_V) \, .
\end{align}
From the above equations we can construct  the following relation
\begin{align} \label{eq:tcZvsFT}
	{\rm BR}(t\to cZ)
=&	 3.5\frac{g^2}{c^2_W}\left[\left( {  5\epsilon\, m_t V_{cb} \over8\sqrt{2} M_*}  \right)^2
	+\left({m_t m_c\over V_{cb} \,\left( M_*\right)^2} \right)^2 \right]  \times
	\frac{{\rm FT}_{\rm mixing}}{ {\rm FT}_t }\left(2\Delta -\frac{ {\rm FT}_{\rm mixing}}{ {\rm FT}_t  } \right) \, ,
\end{align}
where we have used the relevant leading order relation relevant to this case, $m_{t,c} = \epsilon
\lambda^{t,c}_L \lambda^{t,c}_R /2\sqrt{2}M^*_4\,,$ and $\lambda_L^t\sim\lambda_R^t$\,. This correlation is demonstrated in Fig.~\ref{fig:t2czvsFT2} for $M_*=700\,$GeV, $\epsilon=0.3$ and for $0<\theta_V<\pi/4$. 
%

\section{Higgs flavor violation} \label{sec:Htc}

In this section we will investigate flavor violation in the Higgs sector and focus on $t\to ch$. The associated effective Lagrangian can be written as
\begin{align} \label{eq:Linthtc}
{\cal L}^{tch}_{\rm int}=\left(y_{tc,R} \bar t_R c_L+ y_{tc,L} \bar t_L c_R \right) h +h.c. \, .
\end{align}
This interaction leads to the following branching ratio
\begin{align} \label{eq:BRt2ch}
	{\rm BR}(t\ra c  h) \simeq 0.25 (|y_{R,tc}|^2+|y_{L,tc}|^2) \, .
\end{align}

The leading order of the Higgs interaction with fermions in MCHM$_5$ is at $\cO\left(\lambda_{L}\lambda_{R}/M_*^2\right)$~\cite{Agashe:2009di}.
As already discussed above, after intreating out the composite states the effective Lagrangian becomes
\begin{align} \label{eq:massYuk}
	{\cal L}^{h}_{\rm SM}
=	\frac{1}{2\sqrt{2}} \sin\left(\frac{2h}{f}\right) \bar{u}_L^i  \lambda_L^{ik}
	\left[  \left( M_4^{-1}\right)^{kl} - \left( M_1^{-1}\right)^{kl}  \right] (\lambda_R)^{lj} u_R^i\, .
\end{align}
However, at this this order the resulting Yukawa interaction is aligned with the mass matrix, as the spurion structure coincide with that of the mass, see Eq.\eqref{massspurion},
and as a result no flavor violation is generated~\cite{Agashe:2009di}. Thus, Higgs flavor violating couplings appear only at the next order, $\cO\left(\lambda_{L}^2\lambda_{R}^2/
M_*^4\right)$.

We are thus lead to consider higher order terms in the mixings in order to compute the dominant contributions to $t\to hc\,.$
In the holographic Lagrangian the only other terms which are generated by integrating out the composite fermions are the corrections to the kinetic
terms of the elementary fermions
\begin{align}  \label{eq:kin}
{\cal K}_L\simeq
\bar{u}_L \lambda_L U(H) \l(i \sl\d\r)\l[(M_4^\dagger M_4)^{-1}-(M_1^\dagger M_1)^{-1}\r] U(H)^\dagger \lambda_L^\dagger u_L \, ,
\end{align}
where the expression for $u_R$ is achieved by $L\to R$\,. Eqs.~\eqref{eq:massYuk}--\eqref{eq:kin} are the first two terms in the expansion of the form
factor in powers of the external momenta. This expansion is equivalent to a power series in $i\partial\sim m \sim\lambda_{L}\lambda_{R}/M_*$.
Therefore, the operator in Eq.~\eqref{eq:kin} is the only modification to the Yukawa couplings at~$\cO\left(\lambda_{L}^2\lambda_{R}^2/
M_*^4\right)$. By using the equations of motion we can write 
\begin{align}
&{\cal K}_L\simeq  \label{eq:htcOpL}
\bar{u}_L \lambda_L U (H)\l[(M_4^\dagger M_4)^{-1}-(M_1^\dagger M_1)^{-1}\r]U(H)^\dagger \lambda_L^\dagger m^{\rm SM}_{u}u_R \, ,\\
&{\cal K}_R\simeq \label{eq:htcOpR}
\bar{u}_R \lambda_R^\dagger U(H) \l[(M_4 M_4^\dagger)^{-1}-(M_1 M_1^\dagger)^{-1}\r]U(H)^\dagger \lambda_R (m^{\rm SM}_{u})^\dagger u_L \, .
\end{align}
Given that the dominant contributions for $tch$ coupling come from the elements that involve the largest eigenvalue of $m^{\rm SM}_{u}$, i.e. $m_t$,
we can estimate
\begin{align}	
&	y_{tc,L}
\sim  	\frac{\lambda_R^t \lambda_R^c}{M_*^2} \frac{v}{f^2} m_t
\sim 	\frac{m_t m_c}{ f M_*V_{cb}}
\sim 4 \times 10^{-3}\l(\frac{700\,{\rm GeV}}{f}\r)\l(\frac{700\,{\rm GeV}}{M_*}\r)\, , \\
&	y_{tc,R}
\sim 	\frac{\lambda_L^t \lambda_L^c}{M_*^2} \frac{v}{f^2} m_t
\sim	\frac{m_t^2 V_{cb}}{f M_*}
\sim	2 \times 10^{-3}\l(\frac{700\,{\rm GeV}}{f}\r)\l(\frac{700\,{\rm GeV}}{M_*}\r)	\, ,
\end{align}
The resulting branching ratio is
\begin{align}
	{\rm BR}(t\to ch) \sim 5\times 10^{-6} \l(\frac{700\,{\rm GeV}}{f}\r)^2\l(\frac{700\,{\rm GeV}}{M_*}\r)^2 \, .
\end{align}
This is similar to the rate found for the $t\to cZ$ decay mode.
Interesting to note that generically composite models predict that top decay to right handed $c_R$ and Higgs, which is originated by the larger
level of compositeness of $c_R$ compared to $c_L$. As in the case of $t\to cZ$ this can be tested via polarisation measurement of the other-side-top that is predicted to be left handed polarised. The estimated branching ratio is well below the current experimental bounds of ${\rm BR}(t\to
ch)<0.56\%$~\cite{CMS:2014qxa} by CMS and ${\rm BR}(t\to ch)<0.79\%$~\cite{Aad:2014dya} by ATLAS.

\section{$D$ Physics} \label{sec:Dphysics}

In the MCHM$_5$ $Z$ and Higgs mediated flavor violation between the first two generations  is also present in addition to the one between the third and
the second generations. In this section, we estimate these non SM contributions to the $D^0-\overline{D}^0$ mixing as well as to the $D^0\to\ell^+\ell^-$
decay.

Following the above discussion, see Eqs.~\eqref{eq:gZL},~\eqref{eq:gZR} and Appendix~\ref{app:diagM}, the $Zcu$ couplings are estimated as
\begin{align}\label{eq:g_cu}
&	g_{cu,L}
\sim\frac{g}{2\sqrt{2}c_W}\frac{v}{f} \frac{m_t}{M_*}V_{ub}V_{cb}
\sim 3\times10^{-6}\l(\frac{700\,{\rm GeV}}{f}\r)\l(\frac{700\,{\rm GeV}}{M_*}\r) \, , \\
&	g_{cu,R}
\sim 	\frac{g}{2c_W}  \frac{m_cm_u}{M_*^2}\frac{V_{cb}}{V_{ub}}
\sim 6\times10^{-9}\left(\frac{700\,{\rm GeV}}{M_*} \right)^2 \, .
\end{align}
The $hcu$ couplings can be estimated by using Eqs.~\eqref{eq:htcOpL}--\eqref{eq:htcOpR}
\begin{align}
&	y_{cu,L}
\sim 	\frac{m_c^2 m_u}{ f M_* m_t V_{cb}V_{ub}}
\sim 3 \times 10^{-8}\l(\frac{700\,{\rm GeV}}{f}\r)\l(\frac{700\,{\rm GeV}}{M_*}\r)\, , \\
&	y_{cu,R}
\sim	\frac{m_t m_c V_{cb}V_{ub}}{f M_*}
\sim	2 \times 10^{-8}\l(\frac{700\,{\rm GeV}}{f}\r)\l(\frac{700\,{\rm GeV}}{M_*}\r)	\, .
\end{align}

The current experimental bound from $D^0-\overline{D}^0$ mixing on the flavor violating Yukawa was calculated in Ref.~\cite{Harnik:2012pb}. It is
found to be $|y_{cu,L/R}|<7.1\times 10^{-5}$, which is well above the estimated values of $y_{cu,L/R}$.

Regrading the bounds on the $Zcu$ couplings, we follows the procedure given in~\cite{Gedalia:2009kh} with the updated values from~\cite{Amhis:2012bh}. The following upper bounds are found
\begin{align}
&	\mathcal{R}e \left[g_{cu,L}\right] < 6.2 \times 10^{-5}\, , \qquad
	\mathcal{I}m \left[g_{cu,L}\right] < 3.6 \times 10^{-5}\, \\
&	\mathcal{R}e \left[ g_{cu,L}g_{cu,R}\right] < 4.1 \times 10^{-10} \, , \quad
	\mathcal{I}m \left[ g_{cu,L}g_{cu,R}\right] < 1.4 \times 10^{-10} \, .
\end{align}
at one standard deviation. The stronger bound is still an order of magnitude above the estimated effect.

In addition to charm mixing, off-diagonal $Zcu$ coupling leads to non SM decays of $D^0\to \ell^+\ell^-$. The SM long distance contribution to this
branching ratio is estimated as ${\rm BR}(D^0\to \ell^+\ell^-)_{\rm SM}\approx  2.7\times10^{-5}\times{\rm BR}(D^0\to\gamma\gamma)$~
\cite{Burdman:2001tf}. Given the BaBar upper bound of ${\rm BR}(D^0\to\gamma\gamma)<2.2\times 10^{-6}$ at 90\%~CL~\cite{Lees:2011qz}, it is
bounded to be ${\rm BR}(D^0\to \ell^+\ell^-)_{\rm SM}\lesssim 6\times 10^{-11}\,$. By using the result of Ref.~\cite{Golowich:2009ii}, we can
estimate
\begin{align}
	{\rm BR}(D^0\to\ell^+\ell^-)
&	\approx  1.1 \times 10^{-2}\left(g_{cu,L} - g_{cu,R} \right)^2 \nonumber\\
&	\approx 9.9 \times 10^{-14}\l(\frac{700\,{\rm GeV}}{f}\r)^2\l(\frac{700\,{\rm GeV}}{M_*}\r)^2 \, .
\end{align}
The LHCb  95\% CL bound is ${\rm BR}(D^0\to\ell^+\ell^-)<7.6\times 10^{-9}$~\cite{Aaij:2013cza}. We conclude that the effect of non-SM $D^0\to
\ell^+\ell^-$ decays in the MCHM$_5$ models is well below the current experimental bound and the upper bound on the SM prediction.

\section{Conclusions} \label{sec:Conclusions}

In this work we investigate the up flavor structure of composite Higgs models, where we focus on the flavor anarchic minimal $SO(5)$ case. In this
framework the Higgs is a pseudo Nambu-Goldstone boson and can be naturally light. Moreover, the $Zb\bar{b}$ coupling is protected from large non Standard Model contributions due to the custodial symmetry.
We identify the different sources of flavor violation in this framework and emphasise the differences from the anarchic Randall-Sundrum scenario.
The fact that the $SO(5)$ symmetry does not commute with the global flavor group of the model typically leads to reduction in the flavor parameters of the model. We consider the interplay between the fine tuning of the model and flavor violation. We find that generically the tuning of this class of models is worsen in the anarchic case due to the contributions from the additional fermion resonances. Due to the large mixing, they all couple rather strongly to the top sector and, as a result, contribute to the Higgs potential in a ``democratic" manner. We show that, even in the presence of custodial symmetry, large top flavor violating rates are naturally expected.

A $t\to cZ$ branching ratio (BR) of order of $10^{-5}$ is generic for this class of models, and is typically mediated by right-handed currents.
 Thus, polarization measurements of $t\to cZ$ transition provide an additional test for this generic framework. Moreover, the possibility of adopting
charm-tagging at the LHC~\cite{TheATLAScollaboration:2013aia} can help us to distinguish between $t\to cZ$ and $t\to uZ$ transitions (that are further suppressed in our framework).
The above results impliy that this framework can be tested in the next run of the LHC as well as in other
future colliders.

 We also find that the top flavor violation is weakly correlated with an increased amount of fine tuning.
 In the general case the above two phenomena are unrelated. However, in the case in which the misalignment between the composite flavor parameters is
vector like, i.e. the left and right rotations are identical, one can identify a correlation between top flavor violation and the tuning of the model. In a
simplified case, this correlation can be manifested in a simple analytic relation.
Other related flavor violation effects, such as $t \to ch$ and in the $D$
system, are found to be too small to be observed by the current and near future colliders.

\section*{Acknowledgements} \label{sec:Ack}
 We would like to thank R.~Contino for the discussions. GP is supported by the Minerva foundation, the IRG, by the Gruber award, and ERC-2013-CoG grant (TOPCHARM \#\,614794).
\appendix

\section{Diagonalization of the Mass Matrix} \label{app:diagM}

Here we give a more detailed description of the parametric structure of the off diagonal $Z$ couplings. Let us start in the basis where both $\lambda_L$ and $\lambda_R$ are diagonal, where the ratios between their eigenvalues are given in Eq.~\eqref{eq:flavor_anarchy}
\begin{align}
&	\hat{\lambda}_L = \lambda^t_L {\rm diag}\left[ V_{ub},V_{cb}, 1  \right] \sim \lambda^t_L {\rm diag}\left[ \lambda^3,\lambda^2, 1  \right]  \, , \\
&	\hat{\lambda}_R = \lambda^t_R {\rm diag}\left[ m_u/(m_t V_{ub}), m_c/(m_t V_{cb}), 1  \right]\sim \lambda^t_R {\rm diag}\left[\lambda^4, \lambda , 1
\right] \, ,
\end{align}
where $\lambda\sim0.2$ is the Cabibo angle. In this basis both $M_1$ and $M_4$ are anarchic. For convenience we define
\begin{align}
&	\Delta X\equiv M_{4}^{-1} -M_{1}^{-1} \sim 1/ M_*\, , \\
&	 \hat{X}\equiv \left(M^\dagger_{1}M_{1}\right)^{-1} + \left(M^\dagger_{4}M_{4}\right)^{-1}\sim 1/ M_*^2 \, , \\
&	Q \equiv \left( M_{4} M_{4}^\dagger M_{4} \right)^{-1} \sim 1/ M_*^3\, .
\end{align}
The SM particles mass matrix can be written as
\begin{align}
	m^{\rm SM}_u
=&	\frac{\epsilon}{\sqrt{2}}\frac{\lambda_L^t\lambda_R^t}{m_t}  \begin{pmatrix}
	\Delta X_{1,1}m_u  & \Delta X_{1,2} m_c V_{ub}/V_{cb} & \Delta X_{1,3} m_t V_{ub}  \\
	\Delta X_{2,1}m_uV_{cb}/V_{ub}  & \Delta X_{2,2} m_c  & \Delta X_{1,3}m_t V_{cb}  \\
	\Delta X_{3,1}m_u/V_{ub}  & \Delta X_{3,2} m_c/V_{cb}  & \Delta X_{3,3} m_t
	\end{pmatrix}
\sim m_t
	\begin{pmatrix}
	\lambda^7 & \lambda^4 & \lambda^3\\
	\lambda^6 & \lambda^3 & \lambda^2\\
	\lambda^4 & \lambda & 1		
	\end{pmatrix}  \, .
\end{align}
The non-universal parts of the $Z$ couplings matrices are 
\begin{align}
	g^{\rm SM}_{Z,L}
&=	\frac{g\epsilon^2}{4c_W}(\lambda^t_L)^2\begin{pmatrix}
	\hat{X}_{1,1}V_{ub}^2 & \hat{X}_{1,2}V_{cb}V_{ub} & \hat{X}_{1,3}V_{ub} \\
	\hat{X}_{2,1}V_{cb}V_{ub} & \hat{X}_{2,2}V^2_{cb} & \hat{X}_{2,3} V_{cb} \\
	\hat{X}_{3,1}V_{ub} & \hat{X}_{3,2} V_{cb} & \hat{X}_{3,3}
	\end{pmatrix}
	\sim \frac{g\epsilon}{2\sqrt{2}c_W}\frac{m_t}{M_*}
	\begin{pmatrix}
	\lambda^6 & \lambda^5 & \lambda^3\\
	\lambda^5 & \lambda^4 & \lambda^2\\
	\lambda^3 & \lambda^2 & 1		
	\end{pmatrix}  \, , \\
	g^{\rm SM}_{Z,R}
&=	\frac{g\epsilon}{2\sqrt{2}c_W} m^{\rm SM}_u \lambda_L^t\lambda_R^t
	\begin{pmatrix}
	Q_{1,1}m_u  /m_t & Q_{1,2} m_c V_{ub}/(m_tV_{cb})  & Q_{1,3}V_{ub}   \\
	Q_{2,1}m_uV_{cb}/(m_tV_{ub})   & Q_{2,2}m_c  /m_t & Q_{1,3}V_{cb}   \\
	Q_{3,1}m_u/(m_tV_{ub})   & Q_{3,2} m_c/(m_tV_{cb}) &  Q_{3,3}
	\end{pmatrix} +h.c. \nonumber\\
&	\sim  \frac{g}{2c_W} \frac{m_t}{M_*^2} m^{\rm SM}_u
	\begin{pmatrix}
	\lambda^7 & \lambda^4 & \lambda^3\\
	\lambda^6 & \lambda^3 & \lambda^2\\
	\lambda^4 & \lambda & 1		
	\end{pmatrix} +h.c. \, .
\end{align}

The rotation of $m^{\rm SM}_u$ to the mass-eigenstates basis is done by
\begin{align}
	V_{u,L}\sim
	\begin{pmatrix}
		1 & \lambda & \lambda^3 \\
		\lambda & 1 & \lambda^2 \\
		\lambda^3 & \lambda^2  & 1
	\end{pmatrix}  \, , \qquad
	V_{u,R}\sim
	\begin{pmatrix}
		1 & \lambda^3 & \lambda^4 \\
		\lambda^3 & 1 & \lambda \\
		\lambda^4 & \lambda  & 1
	\end{pmatrix} \, .
\end{align}
Since the parametric suppressions of $m^{\rm SM}_u$, $g^{\rm SM}_{Z,L}$ and the relevant part of $g^{\rm SM}_{Z,R}$ are similar, these rotations
will not change the parametric structure of $g^{\rm SM}_{Z,R}$ and $g^{\rm SM}_{Z,L}$. Thus, in the mass-eigenstates basis we can write
\begin{align}
	g^{\rm SM}_{Z,L}
&	\sim \frac{g}{2\sqrt{2}c_W}\epsilon\frac{m_t}{M_*}
	\begin{pmatrix}
	V_{ub}^2 & V_{cb}V_{ub} & V_{ub} \\
	V_{cb}V_{ub} & V^2_{cb} &  V_{cb} \\
	V_{ub} &  V_{cb} & 1
	\end{pmatrix}   \, , \\
	g^{\rm SM}_{Z,R}
&	\sim  \frac{g}{2c_W} \frac{1}{M_*^2}
	\begin{pmatrix}
	m^2_u   &  m_um_c V_{ub}/V_{cb}  & m_um_tV_{ub}   \\
	m_u m_cV_{cb}/V_{ub}   &  m^2_c  & m_c m_tV_{cb}   \\
	m_t m_u/V_{ub}   &  m_t m_c/V_{cb} &  m_t^2
	\end{pmatrix} +h.c. \, .
\end{align}
%



\end{document}